\documentclass[12pt]{iopart}
\pdfoutput=1
\usepackage{graphicx}
\usepackage{dcolumn}
\usepackage{bm}
\usepackage{url}
\usepackage{amssymb}
 \expandafter\let\csname equation*\endcsname\relax 
  \expandafter\let\csname endequation*\endcsname\relax 
\usepackage{amsmath}

\include{epsf}

\newcommand{\R}{{\rm I\! R}}

\newtheorem{Def}{Definition}
\newtheorem{Proposition}[Def]{Proposition}
\newtheorem{Lemma}[Def]{Lemma}

\begin{document}

\title{A semi-Markov model with memory for price changes}
\author{Guglielmo D'Amico}  
\address{Dipartimento di Scienze del Farmaco, Facolt\`a di Farmacia, 
Universit\`a "G. D'Annunzio" di Chieti-Pescara,  66013 Chieti, Italy}
\author{Filippo Petroni}
\address{Dipartimento di Scienze Economiche e Aziendali, Facolt\`a di Economia,
Universit\`a degli studi di Cagliari, 09123 Cagliari, Italy}
\bigskip

\date{\today}
\begin{abstract} 
We study the high frequency price dynamics of traded stocks by a model of returns using a semi-Markov approach.
More precisely we assume that the intraday returns are described by a discrete time homogeneous semi-Markov which depends also on a memory index.  The index is introduced to take into account periods of high and low volatility in the market. 
First of all we derive the equations governing the process and then theoretical results are compared with empirical findings from real data. In particular we analyzed high frequency data from the Italian stock market from first of January 2007 until end of December 2010.  
\end{abstract}
\maketitle
\section{Introduction}
Semi-Markov processes (SMP) are a wide class of stochastic processes which generalize at the same time both Markov chains and renewal processes. The main advantage of SMP is that they allow the use of whatever type of waiting time distribution for modeling the time to have a transition from one state to another one. On the contrary, Markovian models have constraints on the distribution of the waiting times in the states which should be necessarily represented by memoryless distributions (exponential or geometric for continuous and discrete time cases respectively). This major flexibility has a price to pay: the parameters to be estimated are more numerous.\\
\indent Semi-Markov processes generalizes also non-Markovian models based on continuous time random walks used extensively in the econophysics community, see for example \cite{mai00,rab02}.   
SMP have been used to analyze financial data and to describe different problems ranging from credit rating data modeling \cite{dam05} to the pricing of options \cite{dam09,sil04}.

With the financial industry becoming fully computerized, the amount of recorded data, from daily close all the way down to tick-by-tick level, has exploded. Nowadays, such tick-by-tick high-frequency data are readily available for practitioners and researchers alike \cite{gui97,pet03}. It seemed then natural to us trying to verify the semi-Markov hypothesis of returns on high-frequency data, see \cite{dami11}. In that paper we proposed a semi-Markov model showing its ability to reproduce some stylized empirical facts such for example the absence of autocorrelations in returns and the gain/loss asymmetry. In that paper we showed also that the autocorrelation in the square of returns is higher with respect to the Markov model. Unfortunately this autocorrelation was still too small compared to the empirical one. 
 
In order to overcome the problem of low autocorrelation, in this paper we propose an indexed semi-Markov model for price return. More precisely we assume that the intraday returns (up to one minute frequency) are described by a discrete time homogeneous semi-Markov process. We introduce a memory index which takes into account the periods of different volatility in the market. It is well known that the market volatility is autocorrelated, then periods of high (low) volatility may persist for long time. We make the hypothesis that the kernel of the semi-Markov process do depend on which level of volatility the market is at that time.

To check whether our hypothesis describes real data we analyzed high frequency data from the Italian stock market. By Monte Carlo simulations we generated synthetic time series of returns by using the semi-Markov kernel estimated from real data. We showed that the synthetic time series are able to reproduce the autocorrelation of the square of returns as observed for the real time series. We stress the fact that we model returns which are uncorrelated and we obtained correlation in the square of returns. Our approach is different from other models like the GARCH family where the volatility is modeled directly as an autocorrelated process. 

The database used for the analysis is made of high frequency tick-by-tick price data from all the stock in Italian stock market from first of January 2007 until end of December 2010. From prices we then define returns at one minute frequency.

The plan of the paper is as follows. In Section 2 we define the semi-Markov model with memory and we show how to compute its transition probabilities.  In Section 3, we present the empirical results deriving from the application of our model to real stock market data. Finally, in Section 4 we present our conclusion.

\section{The semi-Markov model with memory}
In this section we propose a generalization of the semi-Markov process that is able to represent higher-order dependencies between successive observations of a state variable. One way to increase the memory of the process is by using high-order semi-Markov processes as defined in \cite{limn03}. Here we propose a more parsimonious model having the objective of defining a new model that appropriately describes empirical regularities of financial time series. To this end we extend the model proposed in reference \cite{dami11} allowing the possibility of reproducing long-term dependence in the square of stock returns.

\indent Let $(\Omega,\mathbf{F},P)$ be a probability space and consider the stochastic process
$$
J_{-(m+1)},J_{-m},J_{-(m-1)},...,J_{-1},J_{0},J_{1},...
$$
with a finite state space $E=\{1,2,...,S\}$. In our framework the random variable $J_{n}$ describes the price return process at the $n$-th transition.\\
\indent Let us consider the stochastic process
$$
T_{-(m+1)},T_{-m},T_{-(m-1)},...,T_{-1},T_{0},T_{1},...
$$
with values in $\R$. The random variable $T_{n}$ describes the time in which the $n$-th transition of the price return process occurs.\\
\indent Let us consider also the stochastic process
$$
U_{-(m+1)},U_{-m},U_{-(m-1)},...,U_{-1},U_{0},U_{1},...
$$
with values in $\R$. The random variable $U_{n}$ describes the value of the index process at the $n$-th transition.\\
\indent In reference \cite{dami11b} the process $\{U_{n}\}$ was defined as a reward accumulation process linked to the Markov Renewal Process $\{J_{n},T_{n}\}$. In this paper we introduce a different index process $U_{n}^{m}$ that is defined as follows:
\begin{equation}
\label{funcrela}
U_{n}^{m}=\frac{1}{T_{n}-T_{n-(m+1)}}\sum_{k=0}^{m}\int_{T_{n-1-k}}^{T_{n-k}}f(J_{n-1-k},s)ds,
\end{equation}
where $f:E\times \R \rightarrow \R$ is a Borel measurable bounded function and $U_{-(m+1)}^{m},...,U_{0}^{m}$ are known and non-random.\\
\indent The process $U_{n}^{m}$ can be interpreted as a moving average of the accumulated reward process with the function $f$ as a measure of the rate of reward per unit time.\\
\indent The function $f$ depends on the state of the system $J_{n-1-k}$ and on the time $s$.

As an example you can think of the case in which $m=1$ and $f(J_{n},s)=(J_{n})^{2}$. In this simple case we have that:
\begin{equation}
\label{exa}
U_{n}^{1}=\frac{1}{T_{n}-T_{n-2}}\Bigg((J_{n-1})^{2}\cdot(T_{n}-T_{n-1})+(J_{n-2})^{2}\cdot(T_{n-1}-T_{n-2})\Bigg),
\end{equation}
which expresses a moving average of order $m+1=2$ executed on the series of the square of returns with weights given by the fractions
\begin{equation}
\frac{T_{n}-T_{n-1}}{T_{n}-T_{n-2}};\,\,\,\frac{T_{n-1}-T_{n-2}}{T_{n}-T_{n-2}}.
\end{equation}
It should be noted that the order of the moving average is on the number of transitions. As a consequence, the moving average is executed on time windows of variable length.\\
\indent To construct an indexed model we have to specify a dependence structure between the variables. Toward this end we adopt the following assumption:
\begin{equation}
\label{kernel}
\begin{aligned}
& P[J_{n+1}=j,\: T_{n+1}-T_{n}\leq t |\sigma(J_{h},T_{h},U_{h}^{m}),\, h=-m,...,0,...,n, J_{n}=i, U_{n}^{m}=v]\\
& =P[J_{n+1}=j,\: T_{n+1}-T_{n}\leq t |J_{n}=i, U_{n}^{m}=v]:=Q_{ij}^{m}(v;t),
\end{aligned}
\end{equation}
\noindent where $\sigma(J_{h},T_{h},U_{h}^{m}),\, h\leq n$ is the natural filtration of the three-variate process.\\
\indent The matrix of functions ${\bf Q}^{m}(v;t)=(Q_{ij}^{m}(v;t))_{i,j\in E}$ has a fundamental role in the theory we are going to expose. In recognition of its importance, we call it $\emph{indexed semi-Markov}$ $\emph{kernel}$.\\
\indent The joint process $(J_{n},T_{n})$, which is embedded in the indexed semi-Markov kernel, depends on the moving average process $U_{n}^{m}$, the latter acts as a stochastic index. Moreover, the index process $U_{n}^{m}$ depends on $(J_{n},T_{n})$ through the functional relationship $(\ref{funcrela})$.\\
\indent To describe the behavior of our model at whatever time $t$ we need to define additional stochastic processes.\\
\indent Given the three-dimensional process $\{J_{n}, T_{n}, U_{n}^{m}\}$ and the indexed semi-Markov kernel ${\bf Q}^{m}(v;t)$, we define by
\begin{equation}
\label{stocproc}
\begin{aligned}
& N(t)=\sup\{n\in \mathbb{N}: T_{n}\leq t\};\\
& Z(t)=J_{N(t)};\\
& U^{m}(t)=\frac{1}{t-T_{(N(t)-\theta)-m}}\sum_{k=0}^{m}\int_{T_{(N(t)-\theta)-k}}^{t\wedge T_{(N(t)-\theta)+1-k}}f(J_{(N(t)-\theta)-k},s)ds,
\end{aligned}
\end{equation}
where $T_{N(t)}\leq t < T_{N(t)+1}$ and $\theta =1_{\{t=T_{N(t)}\}}$.\\
\indent The stochastic processes defined in $(\ref{stocproc})$ represent the number of transitions up to time $t$, the state of the system (price return) at time $t$ and the value of the index process (moving average of function of price return) up to $t$, respectively. We refer to $Z(t)$ as an indexed semi-Markov process.\\
\indent The process $U^{m}(t)$ is a generalization of the process $U_{n}^{m}$ where time $t$ can be a transition or a waiting time. It is simple to realize that if $\forall m$, if  $t=T_{n}$ we have that $U^{m}(t)=U_{n}^{m}$.\\  
\indent Let 
$$
p_{ij}^{m}(v):= P[J_{n+1}=j|J_{n}=i,U_{n}^{m}=v].
$$ 
be the transition probabilities of the embedded indexed Markov chain. It denotes the probability that the next transition is in state $j$ given that at current time the process entered in state $i$ and the index process is $v$. It is simple to realize that
\begin{equation}
p_{ij}^{m}(v)=\lim_{t\rightarrow \infty}Q_{ij}^{m}(v;t).
\end{equation}
\indent Let $H_{i}^{m}(v;\cdot)$ be the sojourn time cumulative distribution in state $i\in E$:
\begin{equation}
H_{i}^{m}(v;t):= P[ T_{n+1}-T_{n} \leq t |  J_n=i,\, U_{n}^{m}=v ]= \sum_{j\in E}Q_{ij}^{m}(v;t).
\end{equation}
\indent It expresses the probability to make a transition from state $i$ with sojourn time less or equal to $t$ given the indexed process is $v$.\\  \indent The conditional waiting time distribution function $G$ expresses the following probability:
\begin{equation}
\label{G}
G_{ij}^{m}(v;t):=P[T_{n+1}-T_{n}\leq t \mid J_{n}=i, J_{n+1}=j,U_{n}^{m}=v].
\end{equation}
\indent It is simple to establish that
\begin{eqnarray}
&&G_{ij}^{m}(v;t)=\left\{
                \begin{array}{cl}
                       \ \frac{Q_{ij}^{m}(v;t)}{p_{ij}^{m}(v)}  &\mbox{if $p_{ij}^{m}(v)\neq 0$}\\
                         1  &\mbox{if $p_{ij}^{m}(v)=0$}.\\
                   \end{array}
             \right.
\end{eqnarray}
\indent To properly assess the probabilistic behavior of the system, we introduce the transition probability function:
\begin{equation}
\begin{aligned}
& \phi_{(i_{-(m+1)},i_{-m},...i_{0};j)}^{m}(t_{-(m+1)},t_{-m},...,t_{0};t,V):=\\
& P[Z(t)=j, U^{m}(t)\leq V|J_{0}=i_{0},...,J_{-(m+1)}=i_{-(m+1)},T_{0}=t_{0},...,T_{-(m+1)}=t_{-(m+1)}].
\end{aligned}
\end{equation}
\indent In next proposition we show that the transition probability function of our indexed semi-Markov process satisfies a renewal-type equation. This is an important point because the proposition clearly indicates the mathematical apparatus needed to extend the theoretical investigation of the model. Anyway, the reader can safely skip the proof because the empirical results, illustrated in next section, are obtained via simulation techniques.\\
Before we present a lemma which is a required step in the proof of the proposition.\\
\indent For simplicity of notation set $t_{0}=0$ hereafter.
\begin{Lemma}
\label{Lemma1} 
For all states $j,i_{0},...,i_{-(m+1)}$, for all times $t,t_{0},...,t_{-(m+1)}$ and for any value $V\in \R$ it results that:
\begin{equation}
\label{6a}
\begin{aligned}
& P[Z(t)=j, U^{m}(t)\leq V, T_{1}\leq t|J_{0}=i_{0},...,J_{-(m+1)}=i_{-(m+1)},T_{0}=0,...,T_{-(m+1)}=t_{-(m+1)}]\\
& =E\bigg[1_{\{T_{1}\leq t, J_{N(T_{1})}\in E\}}\cdot\\
& P[Z(t)=j, U^{m}(t)\leq V |T_{1}, J_{N(T_{1})},J_{0}=i_{0},...,J_{-(m+1)}=i_{-(m+1)},T_{0}=0,...,T_{-(m+1)}=t_{-(m+1)}]\bigg].
\end{aligned}
\end{equation}
\end{Lemma}
\textbf{Proof} First of all we remember a property of the conditional expectation. Let $X$ be a random variable and let $A$ be an event, then 
\begin{equation}
\label{proprieta}
E[P[A|X]]=P[A].
\end{equation}
\indent A direct application of this property together with the fact that the events $\{J_{N(T_{1})}=k\}$ are a set of pairwise disjoint events whose union is the entire sample space, produces
\begin{equation*}
\begin{aligned}
& P[Z(t)=j, U^{m}(t)\leq V, T_{1}\leq t|J_{0}=i_{0},...,J_{-(m+1)}=i_{-(m+1)},T_{0}=0,...,T_{-(m+1)}=t_{-(m+1)}]\\
& = E\bigg[P[Z(t)=j, U^{m}(t)\leq V, T_{1}\leq t, J_{N(T_{1})}\in E |T_{1}, J_{N(T_{1})},J_{0}=i_{0},...,J_{-(m+1)}=i_{-(m+1)},\\
& \,\,,T_{0}=0,...,T_{-(m+1)}=t_{-(m+1)}]\bigg]\\
& =E\bigg[1_{\{T_{1}\leq t, J_{N(T_{1})}\in E\}}\cdot\\
& P[Z(t)=j, U^{m}(t)\leq V |T_{1}, J_{N(T_{1})},J_{0}=i_{0},...,J_{-(m+1)}=i_{-(m+1)},T_{0}=0,...,T_{-(m+1)}=t_{-(m+1)}]\bigg].
\end{aligned}
\end{equation*}
\begin{Proposition}
\label{prop}
The probabilities $\phi_{(i_{-(m+1)},i_{-m},...i_{0};j)}^{m}(t_{-(m+1)},t_{-m},...,0;t,V)$ verify the following equation:
\begin{equation}
\label{one}
\begin{aligned}
& \phi_{(i_{-(m+1)},i_{-m},...i_{0};j)}^{m}(t_{-(m+1)},t_{-m},...,0;t,V)\\
& =\delta_{i_{0}j}\Bigg(1-H_{i_{0}}^{m}\big(\frac{1}{-t_{-m-1}}\sum_{k=0}^{m}\int_{t_{-k-1}}^{t_{-k}}f(i_{-k-1},s)ds;t\big)\Bigg)
1_{\Big\{\frac{1}{t-t_{-m}}\sum_{k=1}^{m}\big(\int_{0}^{t}f(i_{0},s)ds+\int_{t_{-k}}^{t_{1-k}}f(i_{-k},s)ds\leq V\big)\Big\}}\\
& +\sum_{s\in E}\int_{0}^{t}Q_{i_{0}\,s}^{m}\bigg( \frac{1}{t-t_{-m-1}}\sum_{k=0}^{m}\int_{t_{-k-1}}^{t_{-k}}f(i_{-k-1},s)ds;d\tau \bigg) \phi_{(i_{-m},...i_{0},s;j)}^{m}(t_{-m},...,0,\tau;t-\tau,V)
\end{aligned}
\end{equation}
\end{Proposition}
\textbf{Proof} First of all let us compute the value of the index process $U^{m}(0)$ given the information set $\{J_{0}=i_{0},...,J_{-(m+1)}=i_{-(m+1)},T_{0}=t_{0},...,T_{-(m+1)}=t_{-(m+1)}\}$. Because $t=0=T_{0}$ is a transition time, we have that $\theta =1$. Moreover $T_{-1-m}=t_{-1-m}$ and $\forall k$ $T_{-k}=t_{-k}<t$. Then, we have that
\begin{equation}
\label{starstar}
U^{m}(0)=\frac{1}{-t_{-m-1}}\sum_{k=0}^{m}\int_{t_{-k-1}}^{t_{-k}}f(i_{-k-1},s)ds.
\end{equation} 
Now, being the events $\{T_{1}>t\}$ and $\{T_{1}\leq t\}$ disjoint, it follows that
\begin{equation}
\label{1}
\begin{aligned}
& P[Z(t)=j, U^{m}(t)\leq V|J_{0}=i_{0},...,J_{-(m+1)}=i_{-(m+1)},T_{0}=0,...,T_{-(m+1)}=t_{-(m+1)}]\\
& =P[Z(t)=j, U^{m}(t)\leq V, T_{1}> t|J_{0}=i_{0},...,J_{-(m+1)}=i_{-(m+1)},T_{0}=0,...,T_{-(m+1)}=t_{-(m+1)}]\\
& +P[Z(t)=j, U^{m}(t)\leq V, T_{1}\leq t|J_{0}=i_{0},...,J_{-(m+1)}=i_{-(m+1)},T_{0}=0,...,T_{-(m+1)}=t_{-(m+1)}].
\end{aligned}
\end{equation}
\indent Observe that
\begin{equation}
\label{2}
\begin{aligned}
& P[Z(t)=j, U^{m}(t)\leq V, T_{1}> t|J_{0}=i_{0},...,J_{-(m+1)}=i_{-(m+1)},T_{0}=0,...,T_{-(m+1)}=t_{-(m+1)}]\\
& = P[T_{1}> t|J_{0}=i_{0},...,J_{-(m+1)}=i_{-(m+1)},T_{0}=0,...,T_{-(m+1)}=t_{-(m+1)}]\\
& \cdot P[Z(t)=j, U^{m}(t)\leq V | T_{1}> t,J_{0}=i_{0},...,J_{-(m+1)}=i_{-(m+1)},T_{0}=0,...,T_{-(m+1)}=t_{-(m+1)}].
\end{aligned}
\end{equation}
\indent The first factor on the right hand side of $(\ref{2})$ is
\begin{equation}
\label{3}
\begin{aligned}
& P[T_{1}> t|J_{0}=i_{0},...,J_{-(m+1)}=i_{-(m+1)},T_{0}=0,...,T_{-(m+1)}=t_{-(m+1)}]=1-H_{i_{0}}(U^{m}(0);t)\\
& =1-H_{i_{0}}^{m}\Big(\frac{1}{-t_{-m-1}}\sum_{k=0}^{m}\int_{t_{-k-1}}^{t_{-k}}f(i_{-k-1},s)ds;t\Big).
\end{aligned}
\end{equation}
\indent The second factor on the right hand side of $(\ref{2})$ is
\begin{equation}
\label{4}
\begin{aligned}
& P[Z(t)=j, U^{m}(t)\leq V | T_{1}> t,J_{0}=i_{0},...,J_{-(m+1)}=i_{-(m+1)},T_{0}=0,...,T_{-(m+1)}=t_{-(m+1)}]\\
& = P[U^{m}(t)\leq V |Z(t)=j, T_{1}> t,J_{0}=i_{0},...,J_{-(m+1)}=i_{-(m+1)},T_{0}=0,...,T_{-(m+1)}=t_{-(m+1)}]\\
& \cdot P[Z(t)=j| T_{1}> t,J_{0}=i_{0},...,J_{-(m+1)}=i_{-(m+1)},T_{0}=0,...,T_{-(m+1)}=t_{-(m+1)}].
\end{aligned}
\end{equation}
\indent Now, observe that, $T_{1}>t$ means that the time of next transition exceeds $t$ and, therefore, up to $t$ the process remains in state $i_{0}$. Consequently,  
$$
P[Z(t)=j| T_{1}> t,J_{0}=i_{0},...,J_{-(m+1)}=i_{-(m+1)},T_{0}=0,...,T_{-(m+1)}=t_{-(m+1)}]=\delta_{i_{0}j}.
$$ 
Moreover, because $t\neq T_{N(t)}$ we have that $\theta =0$ and consequently 
$$
U^{m}(t)=\frac{1}{t-T_{N(t)-m}}\sum_{k=0}^{m}\int_{T_{N(t)-k}}^{t\wedge T_{N(t)+1-k}}f(J_{N(t)-k},s)ds.
$$
Since $T_{1}>t$, we have that $T_{N(t)}=T_{0}=0$ and by substitution we get in $U^{m}(t)=\frac{1}{t-T_{-m}}\sum_{k=0}^{m}\int_{T_{-k}}^{t\wedge T_{1-k}}f(J_{-k},s)ds$. Finally, $T_{-k}=t_{-k}$ and $J_{-k}=i_{-k}$, for all $k$, produces
\begin{equation}
\label{diciassette}
U^{m}(t)=\frac{1}{t-t_{-m}}\sum_{k=1}^{m}\Big(\int_{0}^{t}f(i_{0},s)ds + \int_{t_{-k}}^{t\wedge t_{1-k}}f(i_{-k},s)ds\Big).
\end{equation} 
\indent In formula $(\ref{diciassette})$ we computed the value of the $U^{m}(t)$ process given the information set $\{T_{1}> t,J_{0}=i_{0},...,J_{-(m+1)}=i_{-(m+1)},T_{0}=0,...,T_{-(m+1)}=t_{-(m+1)}\}$; this proves that the random variable $U^{m}(t)$ is $\sigma(J_{0},...,J_{-(m+1)},T_{1},T_{0},...,T_{-(m+1)})$-measurable.\\
\indent By using the properties of the indicator function we have that: 
\begin{equation}
\label{5}
\begin{aligned}
& P[U^{m}(t)\leq V |Z(t)=j, T_{1}> t,J_{0}=i_{0},...,J_{-(m+1)}=i_{-(m+1)},T_{0}=0,...,T_{-(m+1)}=t_{-(m+1)}]\\
& =E[1_{\{U^{m}(t)\leq V\}}|Z(t)=j, T_{1}> t,J_{0}=i_{0},...,J_{-(m+1)}=i_{-(m+1)},T_{0}=0,...,T_{-(m+1)}=t_{-(m+1)}].
\end{aligned}
\end{equation}
\indent To evaluate the expectation in formula $(\ref{5})$, note again that the conditions $(Z(t)=j,T_{1}>t,J_{0}=i_{0})$ are only compatible for $Z(t)=i_{0}$, then the value of the process $U^{m}(t)$ is that expressed in formula $(\ref{diciassette})$. Then we can express $(\ref{5})$ as follows:
\begin{equation}
\label{5bis}
\begin{aligned}
=& E\bigg[1_{\{U^{m}(t)\leq V\}}|Z(t)=j, T_{1}> t,J_{0}=i_{0},...,J_{-(m+1)}=i_{-(m+1)},T_{0}=0,...,T_{-(m+1)}=t_{-(m+1)}\\
& ,U^{m}(t)=\frac{1}{t-t_{-m}}\sum_{k=1}^{m}\Big(\int_{0}^{t}f(i_{0},s)ds + \int_{t_{-k}}^{t\wedge t_{1-k}}f(i_{-k},s)ds\Big)\bigg]\\
& =1_{\big\{\frac{1}{t-t_{-m}}\sum_{k=1}^{m}\big(\int_{0}^{t}f(i_{0},s)ds + \int_{t_{-k}}^{t\wedge t_{1-k}}f(i_{-k},s)ds\big)\leq V \big\}}.
\end{aligned}
\end{equation}
\indent Now let us consider the computation of the second addend on the right hand side of formula $(\ref{1})$. By Lemma $(\ref{Lemma1})$ we know that
\begin{equation}
\label{6}
\begin{aligned}
& P[Z(t)=j, U^{m}(t)\leq V, T_{1}\leq t|J_{0}=i_{0},...,J_{-(m+1)}=i_{-(m+1)},T_{0}=0,...,T_{-(m+1)}=t_{-(m+1)}]\\
& =E\big[1_{\{T_{1}\leq t, J_{N(T_{1})}\in E\}}\cdot\\
& P[Z(t)=j, U^{m}(t)\leq V |T_{1}, J_{N(T_{1})},J_{0}=i_{0},...,J_{-(m+1)}=i_{-(m+1)},T_{0}=0,...,T_{-(m+1)}=t_{-(m+1)}]\big].
\end{aligned}
\end{equation}
Because the random variable $U^{m}(0)$ is $\sigma(J_{0},...,J_{-(m+1)},T_{0},...,T_{-(m+1)})$-measurable, see formula $(\ref{starstar})$, then formula $(\ref{6})$ is equivalent to formula $(\ref{6bis})$
\begin{equation}
\label{6bis}
\begin{aligned}
& =E\bigg[1_{\{T_{1}\leq t, J_{N(T_{1})}\in E\}}P[Z(t)=j, U^{m}(t)\leq V |T_{1}, J_{N(T_{1})},\\
& U^{m}(0)=\frac{1}{-t_{-m-1}}\sum_{k=0}^{m}\int_{t_{-k-1}}^{t_{-k}}f(i_{-k-1},s)ds,\\
& J_{0}=i_{0},...,J_{-(m+1)}=i_{-(m+1)},T_{1},T_{0}=0,...,T_{-(m+1)}=t_{-(m+1)}]\bigg].
\end{aligned}
\end{equation}
\indent Now observe that 
\begin{equation}
\begin{aligned}
& P\big[T_{1}\in (\tau, \tau +d\tau ), J_{N(T_{1})}=s|U^{m}(0)=\frac{1}{-t_{-m-1}}\sum_{k=0}^{m}\int_{t_{-k-1}}^{t_{-k}}f(i_{-k-1},s)ds,\\
& J_{0}=i_{0},...,J_{-(m+1)}=i_{-(m+1)},T_{0}=0,...,T_{-(m+1)}=t_{-(m+1)}\big]\\
& = P[T_{1}\in (\tau, \tau +d\tau ), J_{N(T_{1})}=s|U^{m}(0)=\frac{1}{-t_{-m-1}}\sum_{k=0}^{m}\int_{t_{-k-1}}^{t_{-k}}f(i_{-k-1},s)ds,J_{0}=i_{0}]\\
& = Q_{i_{0}\,s}^{m}\bigg(\frac{1}{-t_{-m-1}}\sum_{k=0}^{m}\int_{t_{-k-1}}^{t_{-k}}f(i_{-k-1},s)ds; d\tau \bigg).
\end{aligned}
\end{equation}
\indent Moreover notice that
\begin{equation}
\label{7}
\begin{aligned}
& P[Z(t)=j, U^{m}(t)\leq V|T_{1}, J_{N(T_{1})},J_{0}=i_{0},...,J_{-(m+1)}=i_{-(m+1)},T_{0}=0,...,T_{-(m+1)}=t_{-(m+1)}]\\
& =P[Z(t)=j, U^{m}(t)\leq V|T_{1}, J_{N(T_{1})},J_{0}=i_{0},...,J_{-m}=i_{-m},T_{0}=0,...,T_{-m}=t_{-m}]\\
& =\phi_{(i_{-m},...,i_{0},J_{N(T_{1})};j)}^{m}(t_{-m},...,0,T_{1} ;t-T_{1},V).
\end{aligned}
\end{equation}
\indent Then, after an integration over the possible values of $T_{1}\in (0,t]$ and a summation over the values of $J_{N(T_{1})}\in E$, formula $(\ref{6})$ becomes equal to
\begin{equation} 
\sum_{s\in E}\int_{0}^{t}Q_{i_{0}\,s}^{m}(U^{m}(0);d\tau)\phi_{(i_{-m},...,i_{0},s;j)}^{m}(t_{-m},...,0,\tau ;t-\tau,V).
\end{equation}
\indent A substitution in $(\ref{1})$ completes the proof.\\
\indent Proposition $\ref{prop}$ gives a renewal-type equation for transition probabilities and represents a generalization of the evolution equation for transition probabilities of the semi-Markov process. Equation $(\ref{one})$ could be used for asymptotic analysis in order to obtain explicit representation of the autocorrelation function. Moreover the numerical solution of equation $(\ref{one})$, based on quadrature formulas, could represent an alternative to the simulative approach used in the next section.

\section{Empirical results}

To check the validity of our model we perform a comparison of the behavior of real data returns
and returns generated through Monte Carlo simulations based on the model.
In this section we describe the database of real data used for the analysis, the method used to simulate
synthetic returns time series and, at the end, we compare results from real and simulated data.   

\subsection{Database description}
The data we used in this work are tick-by-tick quotes of indexes and stocks downloaded 
from $www.borsaitaliana.it$ for the period January 2007-December 2010 (4 full years). 
The data have been re-sampled to have 1 minute frequency. Consider a single day (say day $k$ with $1 \le k \le d$)  
where $d$ is number of traded days in the time series. In our case we consider 
four years of trading (from the first of 
January 2007 corresponding to $d=1076$).
The market in Italy fixes the opening price at a
random time in the first minute after 9 am,
continuous trading starts immediately after and ends just before  5.25 pm,
finally the closing price is fixed just after 5.30 pm.
Therefore, let us define $S(t)$ as the price of the last trading
before 9.01.00 am , $S(t+1)$ as the price of the last trading
before 9.02.00 am and so on until
$S(nk)$ as the price of the last trading
before 5.25.00  pm.  
If there are no transactions in the minute,
the price remains unchanged
(even in the case the title is suspended and reopened
in the same day).
Also define
$S(nk+1)$ as the opening price and $S(nk)$ as the closing price.
With this choice $n=507$.
There was a small difference before the 28th of September 2009 since
continuous trading started at 9,05 am, and therefore
prior of that date we have $n=502$.
Finally, if the title has a delay in opening
or it closes in advance (suspended but not reopened),
only the effective trading minutes 
are taken into account. In this case $n$ will be smaller than 507.
The number of returns analyzed is then roughly 508000 for each stock.
We analyzed all the stocks in the FTSEMIB which are the 40 most capitalized
stocks in the Italian stock market.

\begin{figure}
\centering
\includegraphics[width=8cm]{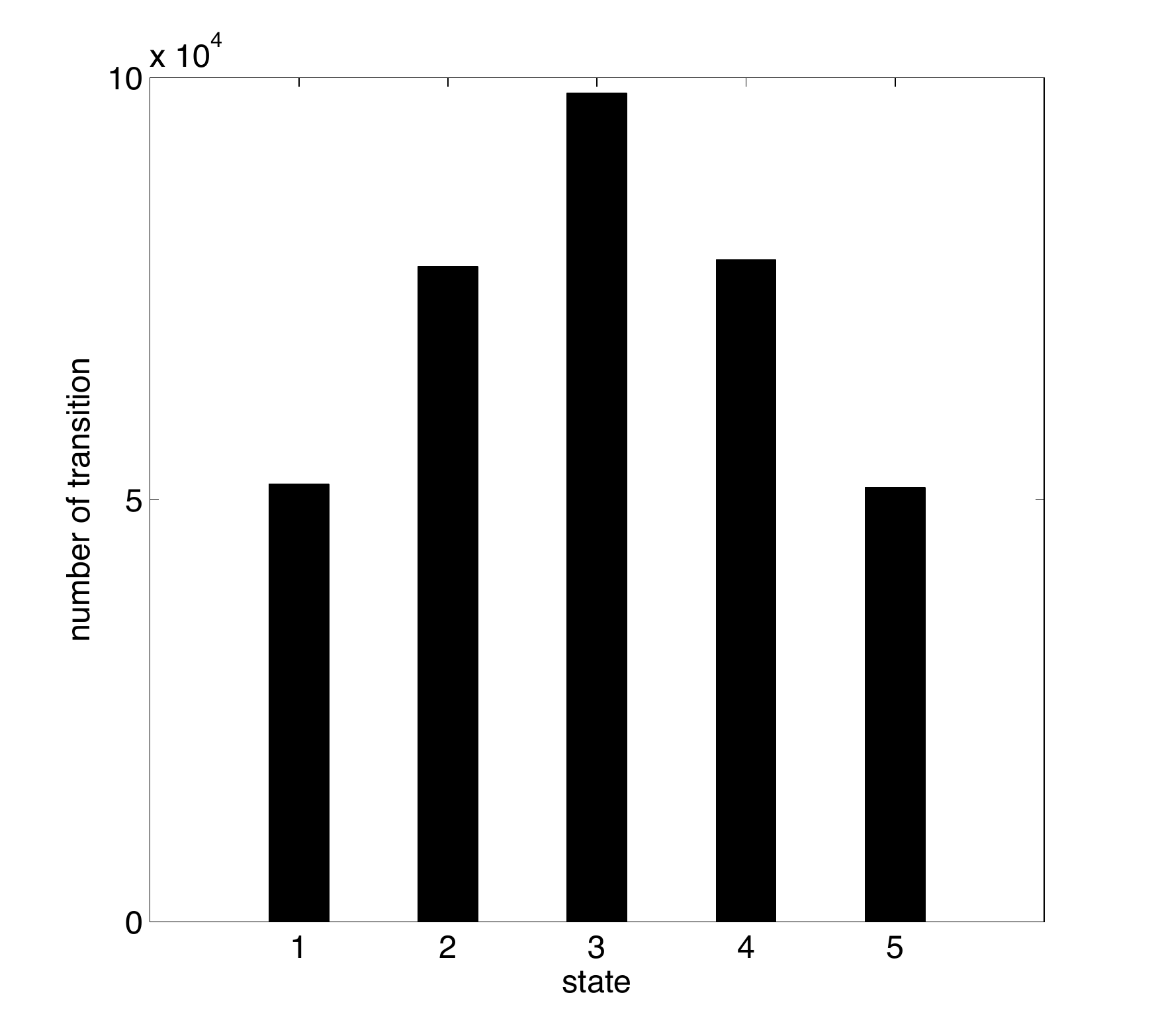}
\caption{Number of transitions for the embedded Markov chain} \label{tra}
\end{figure}

To be able to model returns as a semi-Markov process the state space has to be discretized.
In the example shown in this work we discretized returns into 5 states chosen to be symmetrical with respect to returns equal zero. Returns are in fact already discretized in real data due to the discretization of stock prices which is fixed by each stock exchange and depends on the value of the stock. Just to make an example, in the Italian stock market for stocks with value between 5.0001 and 10 euros the minimum variation is fixed to 0.005 euros (usually called tick). We then tried to remain as much as possible close to this discretization. In Figure \ref{tra} we show an example of the number of transition from state $i$
to all other states for the embedded Markov chain.

\subsection{Monte Carlo simulations}
In the model described in the previous section and in particular in the definition of the index process $U^m$ the function $f:E\times \R \rightarrow \R$ is any Borel measurable bounded function. To perform simulations, we choose a function which is both motivated by simplicity (we want to keep the model as simple as possible) and by real market behavior. 

Let us briefly remind that volatility of real market is long range positively autocorrelated and then clustered in time. This implies that, in the stock market, there are periods of high and low volatility. Motivated by this empirical facts we suppose that also the transition probabilities depends on whether the market is in a high volatility period or in a low one. We then fixed the function $f$ to be the square of returns and the index $U^m$ to be a moving average of this values as described in the example given in eq. \ref{exa} but with different values of the memory $m$. Note that the memory is the number of transitions. The index $U^m$ obtained from the given definition of $f$ was also discretized into 5 states of low, medium low, medium, medium high and high volatility.
 
According to these choices we estimated, from real data, the probabilities $Q_{ij}^m(v;t)$ defined in formula (\ref{kernel}) for different values of $m$. For the results shown below $m$ was chosen to run from 5 to 200 transitions with a step of 5.

Then, this probabilities have been used to simulate synthetic time series of returns needed to compare results from real data and the model as described in the next section. Note that these are step-by-step simulations in which the index $U^m$ has to be calculated from the last $m$ simulated transitions.

\subsection{Results on autocorrelation function}
A very important feature of stock market data is that, while returns are uncorrelated and show an i.i.d. like behavior, their square or absolute values are long range correlated. It is very important that theoretical models of returns do reproduce this features.   
We then tested our model to check whether it is able to reproduce such behavior. 
Given the presence of the parameter $m$ in the index function, we also tested the autocorrelation behavior as a function of m.

If $R$ indicates returns, the time lagged $(\tau)$ autocorrelation of the square of returns is defined as 

\begin{equation}
\label{autosquare}
\Sigma(\tau)=\frac{Cov(R^2(t+\tau),R^2(t))}{Var(R^2(t))}
\end{equation}

\begin{figure}
\centering
\includegraphics[height=8cm]{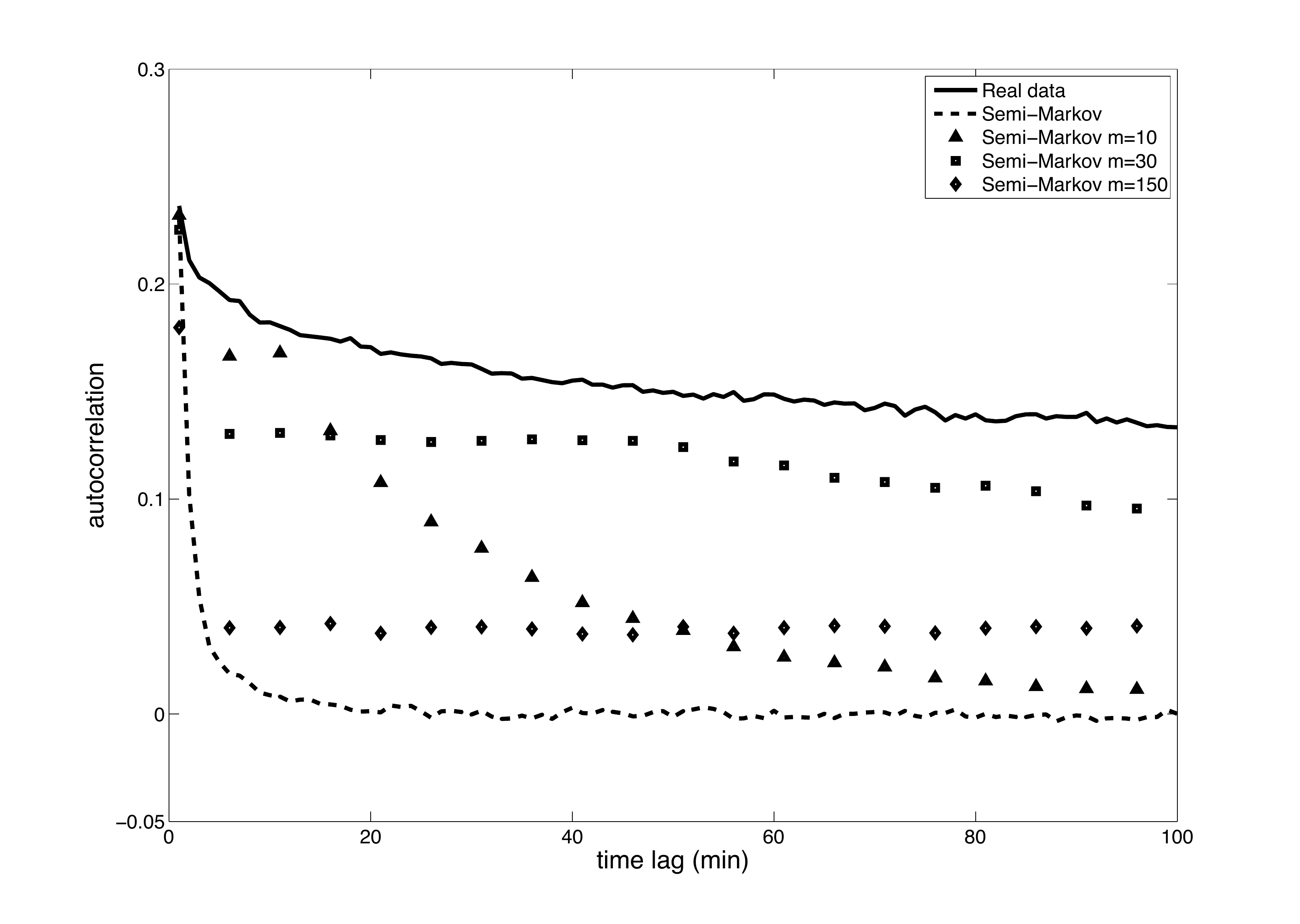}
\caption{Autocorrelation functions of real data (solid line) and of 4 synthetic time series as described in the label.}\label{fig1}
\end{figure}

We estimated $\Sigma(\tau)$ for real data and for returns time series simulated with different values of the memory $m$. 
The time lag $\tau$ was made to run from 1 minute up to 100 minutes. Note that to be able to compare results for $\Sigma(\tau)$ each
simulated time series was generated with the same length as real data.
Results for few values of $m$, for real data and for a semi-Markov model without index are shown in Figure \ref{fig1}.

\begin{figure}
\centering
\includegraphics[height=8cm]{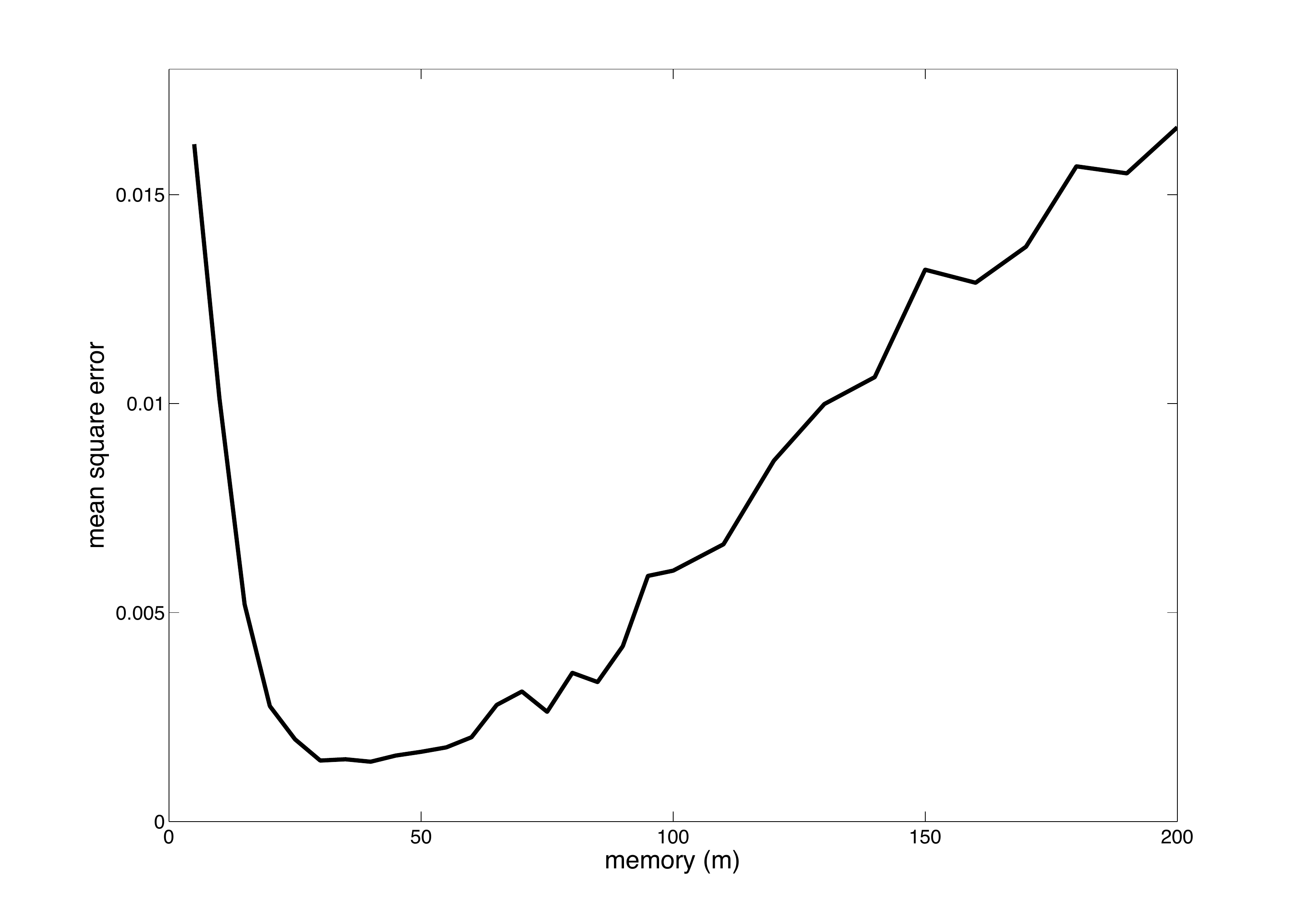}
\caption{Mean square error between autocorrelation function of real data and synthetic data as a function of the memory value $m$.}\label{fig2}
\end{figure}

As expected, real data do show a long range correlation of volatility let us than analyze results for the synthetic time series. The simple semi-Markov model starts at the same value but the persistence is very short and after few time steps the autocorrelation decrease to zero. A very interesting behavior is instead shown by the semi-Markov models with memory index. If a small memory ($m=10$ in the shown example) is used, the autocorrelation is already persistent but again decreases faster than real data. With a longer memory ($m=30$) the autocorrelation remain high for a very  long period and also its value is very close to that of real data. If $m$ is increased further the autocorrelation drops again to small values.
This behavior suggest the existence of an optimal memory $m$. In our opinion one can justify this behavior by saying that short memories are not enough to identify in which volatility status is the market, too long memories mix together different status and then much of the information is lost in the average. 
All this is shown in Figure \ref{fig2} where the mean square error between each autocorrelation function of simulated time series and the autocorrelation function of the real data as a function of $m$ is computed. It can be noticed that there exist an optimal value of the memory $m$ that makes the autocorrelation of simulated data closer to that of real data. 

\section{Conclusions}
We have modeled financial price changes through a semi-Markov model where we have added a memory index. Our work is motivated by two main results: the existence in the market of periods of low and high volatility and our previous work \cite{dami11}, where we showed that the semi-Markov model, even if more realistic than a simple Markov model, is not able to capture all the correlation in the square of returns present in real data.
The results presented here show that the semi-Markov kernel is influenced by the past volatility. In fact, if the past volatility is used as a memory index, the model is able to reproduce quite well the behavior of market returns: the returns generated by the model are uncorrelated while the square of returns present a long range correlation very similar to that of real data. 

We have also shown that the time length of the memory does play a crucial role in reproducing the right autocorrelation's persistence indicating the existence of an optimal value.

We stress that out model is very different from those of the ARCH/GARCH family. We do not model directly the volatility as a correlated process. We model returns and by considering the semi-Markov kernel conditioned by a memory index the volatility correlation comes out freely.

\section*{Acknowledgments}
We are very grateful to Maurizio Serva, Raimondo Manca and Giuseppe Di Biase for helpful discussion. We also thank an anonymous referee for his precious comments.
\newpage

\end{document}